\documentstyle[12pt]{article}     
\font\smallit= cmti9 scaled\magstep2
\def\matrix#1#2{\left[ \begin{array}{*{#1}{c}} #2 \end{array} \right]}
\def\rtwo{\frac1{\sqrt2}}
\def\eref#1{(\ref{#1})}

\begin{document}
\thispagestyle{empty}
\pagestyle{plain}
\date{December, 1997}
\begin{titlepage}
\bigskip
\hfill IASSNS-HEP-97-133\\
\phantom{}\hfill KEK Preprint 97-232\\
\bigskip\bigskip\bigskip\bigskip

\centerline{\Large\bf BPS Saturation from Null Reduction}

\bigskip\bigskip
\bigskip\bigskip

\centerline{\large Reinhold W. Gebert\dag\ 
\footnote{\tt gebert@ias.edu}
and Shun'ya Mizoguchi\ddag\ 
\footnote{\tt mizoguch@tanashi.kek.jp}
}
\setcounter{footnote}{0}
\medskip
\vskip 5mm
\centerline{
\dag\ \smallit Institute for Advanced Study, School of Natural Science}
\centerline{ 
\smallit Olden Lane, Princeton, NJ 08540, U.S.A.
}
\medskip
\centerline{
\ddag\ \smallit Institute of Particle and Nuclear Studies, KEK}
\centerline{\smallit Tanashi, Tokyo 188 Japan}
\bigskip\bigskip

\begin{abstract}
\baselineskip=16pt
We show that any $d$-dimensional strictly stationary, asymptotically 
Minkowskian solution $(d\ge 4)$ of a null reduction of $d+1$-dimensional 
pure gravity must saturate the BPS bound provided that the KK vector 
field can be identified appropriately. We also argue that it is 
consistent with the field equations.
\end{abstract}
\end{titlepage}

BPS solutions in supergravity theories play an important
role in probing non-perturbative features of M theory and string
theory. Some solutions are known to be obtained by
infinitely boosting a static solution along some compactifed
direction. For example, let us consider the $d=10$ Schwarzschild
solution smeared along the eleventh direction:
\begin{equation}
d s_{11}^2
=
-\left(1-\frac{\mu}{r^7}\right) d t^2
+\left(1-\frac{\mu}{r^7}\right)^{-1} d r^2
+r^2 d\Omega_8 +d y^2,
\end{equation}
where $r^2=x_1^2+\cdots+x_9^2$, which is a solution of $d=11$
supergravity (with the three-from set to zero). The global Lorentz
transformation
$t \rightarrow  t\cosh\beta - y\sinh\beta$, 
$y \rightarrow - t\sinh\beta+ y\cosh\beta$
boosts the solution along the $y$-direction. In the limit
$\beta\rightarrow\infty$, $\mu\rightarrow0$, $\mu
{\rm e}^{2\beta}\rightarrow 4Q$, this becomes
\begin{equation}
d s_{11}^2=-d t^2+d y^2+W(d t-d y)^2 + d r^2 + r^2d\Omega_8,
\end{equation}
where $W=Q/r^7$. Compactifying $y$, one reads off a solution
of type IIA supergravity:
\begin{eqnarray}
d s_{10A}^2
&=& -K^{-1/2}d t^2+K^{1/2}(d r^2+r^2d\Omega_8), \nonumber\\
{\rm e}^{2\phi}
&=&K^{3/2},\\
A_\mu
&=&-\delta_{\mu}^t WK^{-1},  \nonumber
\end{eqnarray}
with $K=1+W$. This is nothing but the $d=10$ extremal 0-brane solution
\cite{HorStr91} expressed in the isotropic radial coordinate 
$\tilde{r}^7 = r^7+Q$. 
More complicated examples can be found in \cite{RusTse97}. 
The BPS bound $M\ge c|Q|$ (with $c$ being a positive 
convention-dependent constant) is saturated by a
Kaluza--Klein (KK) electric charge in the simplest cases, while other
charged solutions can be obtained by duality symmetries.

The BPS saturation thus achieved can be intuitively understood
in the following way. Suppose that we are given a static solution with
energy-momentum $(d+1)$-vector $(E,P_\perp,P_\parallel)=(M',0,0)$. The
longitudinal momentum $P_\parallel$ increases as we boost it, and in
the infinite-boost limit the energy-momentum approaches
$(M,0,M)$ for some $M$. If we compactify the longitudinal direction,
$P_\parallel$ becomes the KK charge, so that the solution
can be viewed as a static BPS solution with mass and charge being
equal.

Any infinitely boosted solution of this kind necessarily possesses a
null Killing vector field, and hence is a solution of a null reduction
of a higher-dimensional theory. In this letter we show that, in arbitrary 
dimensions $d\ge 4$, any strictly stationary, asymptotically Minkowskian 
solution of a null reduction of $d+1$-dimensional pure gravity must 
be a BPS solution\footnote{or at least an extremal solution in some 
sense; anticipating an application to supersymmetric theories 
we will employ the former terminology throughout the paper.}
provided that the KK vector field can be identified 
appropriately. In dimensions where the $d$-dimensional theory can be 
obtained as a bosonic sector of $N=2$ supergravity, the KK charge becomes 
a central charge of the superalgebra and the solution allows a Killing 
spinor.

Let us consider a pseudo-Riemannian manifold admitting a pair of
commuting Killing vector fields one of which is assumed to be null. We
start with the following parameterization of the vielbein (see
\cite{JulNic95} for the general framework of null reduction):
\begin{equation}
{E_M}^{\tilde{A}} =
\matrix3{ 
{E_m}^a & u_m & SC_m \\
0     & u_w & SC_w \\
0     & 0   & S }
\end{equation}
with  a flat lightcone metric
\begin{equation}
\eta_{\tilde{A}\tilde{B}} =
\matrix3{
\delta_{ab} & 0 & 0 \\
0           & 0 & 1 \\
0           & 1 & 0 }.
\end{equation}
We shall use the following conventions. Capital letters $M,N,\ldots$
and $\tilde{A},\tilde{B},\ldots$ (as well as $A,B,\ldots$ below) will
denote curved and flat indices, respectively, in $d+1$
dimensions. Upon dimensional reduction they are split into transversal
and longitudinal indices, i.e., $M=(m,w,v)$ and $\tilde{A}=(a,+,-)$
where $a,b,\ldots,m,n,\ldots=1,\ldots,d-1$. By the above
choice of parameterization the local $SO(d,1)$ Lorentz invariance is
broken down to $SO(d-1)\times SO(1,1)$. This is in contrast to
dimensional reduction with a null Killing vector alone \cite{JulNic95} 
for which the residual tangent space symmetry is the inhomogeneous
Lorentz group $ISO(d-1)$.

The  $wv$ part of the metric reads
\begin{equation}
\matrix2{
G_{ww} & G_{wv} \\
G_{vw} & G_{vv} }
=
\matrix2{
2Su_wC_w & Su_w \\
Su_w     & 0 }.
\end{equation}
The two Killing vectors corresponding to the dimensional reduction are
taken to have components $\omega^M=(0,\ldots,0,1,0)$ and
$\xi^M=(0,\ldots,0,1)$, so that
\begin{equation}
\omega\equiv\omega^M\partial_M=\partial_w, \qquad
\xi\equiv\xi^M\partial_M=\partial_v,
\end{equation}
respectively, and $\xi$ is indeed null.

To identify the KK vector field we change tangent space
lightcone coordinates $\tilde{A}=(a,+,-)$ into standard Minkowski
coordinates $A=(a,0,d)$ by applying a similarity transformation to
$\eta_{\tilde{A}\tilde{B}}$ so that it becomes
\begin{equation}
\eta_{AB} =
\matrix3{
\delta_{ab} & 0  & 0 \\
0           & -1 & 0 \\
0           & 0  & 1 }.
\end{equation}
Consequently, the vielbein changes into
\begin{equation}
{E_M}^A =
\matrix3{
{E_m}^a & \rtwo(u_m-SC_m) & \rtwo(u_m+SC_m) \\
0       & \rtwo(u_w-SC_w) & \rtwo(u_w+SC_w) \\
0       & -\rtwo S         & \rtwo S }.
\end{equation}
To make contact with ordinary dimensional reduction w.r.t.\ a
spacelike Killing vector $\partial_x$ let us now perform a 
change of coordinates 
$w = pt+rx$,
$v = qt+sx$
with some constants $p,q,r,s$ such that $\Delta := ps-rq \neq
0$. Taking also the freedom of a local $SO(1,1)$ transformation into
account, the $tx$ part of the vielbein becomes
\begin{eqnarray}
\matrix2{
{E_t}^0 & {E_t}^d \\
{E_x}^0 & {E_x}^d } 
&=&
\matrix2{ p & q \\ r & s }
\rtwo\matrix2{
u_w-SC_w & u_w+SC_w \\
- S         &  S }
\matrix2{
\cosh\theta & \sinh\theta \\
\sinh\theta & \cosh\theta } \nonumber \\
&=&
\rtwo\matrix2{
pu_we^\theta-(pC_w+q)S{\rm e}^{-\theta} &
pu_we^\theta+(pC_w+q)S{\rm e}^{-\theta} \\
ru_we^\theta-(rC_w+s)S{\rm e}^{-\theta} &
ru_we^\theta+(rC_w+s)S{\rm e}^{-\theta} } \nonumber \\
&=&
\matrix2{
\Delta Su_w/\rho & \rho p/r - \Delta Su_w/\rho \\
     0   & \rho }, \label{txpart}
\end{eqnarray}
where $\rho:= r[2Su_w(C_w+s/r)]^{1/2}$ and we have chosen $\theta$ 
such that ${\rm e}^{2\theta}=(rC_w+s)S/ru_w$. 
The constants $r,s$ must satisfy 
$s/r>-\inf C_w$ because $\theta$ is real\footnote{Note 
that $u_m$, $u_w$ and 
$S$ appear only through $Su_m$ or $Su_w$ in the ($d+1$)-metric.  
Since $\det{E_M}^A=Su_w \det{E_m}^a $, we may assume $S>0$ 
and $u_w>0$.}. 
Besides we take $r>0$ so that we may 
identify $\rho$ with (the exponential of) the dilaton.  
The $tx$ part of the $(d+1)$-metric then reads
\begin{equation}
\matrix2{
G_{tt} & G_{tx} \\
G_{xt} & G_{xx} }
=
\matrix2{
2pu_wS(pC_w+q) & \rho^2p/r-\Delta Su_w\\
\rho^2p/r-\Delta Su_w  & \rho^2 }.
\end{equation}
We observe that because of $\rho^2>0$ the Killing vector $\partial_x=
r\partial_w + s\partial_v$ is always spacelike, while $\partial_t=
p\partial_w + q\partial_v$ becomes timelike if $p\neq0$, $q/p<-\sup C_w$. 
We have thus obtained a stationary configuration in $d$ 
dimensions from the null reduction with an extra Killing 
vector field. Any system of two commuting Killing vectors one of which 
is null can be viewed in this way as consisting of a spacelike and a 
timelike Killing vector. Note, 
however, that the converse is not correct in general.

To identify the physical fields we equate
\begin{eqnarray}
{E_M}^A
=
\matrix3{
{E_m}^a & \rtwo(u_m-SC_m) & \rtwo(u_m+SC_m) \\
0       & \Delta Su_w/\rho        & \rho p/r - \Delta Su_w/\rho\\
0       &         0       & \rho } ~=~
\matrix2{
\rho^{\chi}{e_\mu}^\alpha & \rho A_\mu \\
          0                             & \rho }, 
\label{identification}
\end{eqnarray}
where $\mu=(m,t)$ and $\alpha=1,\ldots,d-1,0$. 
There is an arbitrariness specified by $\chi$ 
in splitting the dilaton factor from the $d$-metric 
${e_\mu}^\alpha$. If one takes $\chi=-\frac{1}{d-2}$, the $eR$ term takes  
the canonical form without any dilaton factor in the reduced action.
Another common choice is $\chi=-1/2$ in $d=10$, 
known as the ``string metric'', in which the dilaton factor 
disappears from the $F^2$ term. In our case 
yet another choice turns out to be more convenient, as we will see below.

From \eref{identification} we find the following expressions for 
the dilaton and the components of the KK vector field, respectively,
\begin{equation}
{\rm e}^{2\phi}:=\rho,\qquad
A_m= \rtwo(u_m+SC_m)\rho^{-1}, \qquad
A_t= \frac p r - \frac\Delta{\rho^2} Su_w.
\end{equation}
In addition, adopting $\chi=1$, we get the condition
\begin{equation}
{e_t}^0= \frac p r - A_t.
\label{A-Relation} 
\end{equation}
The BPS saturation for strictly stationary, asymptotically Minkowskian 
solutions is an immediate consequence of the relation \eref{A-Relation}. 
More precisely, we assume that $x$ can be globally separated 
from the $d+1$-dimensional spacetime with a suitable identification 
of different local neighborhoods. We then require that the 
$d$-bein ${e_\mu}^\alpha$ in a local coordinate of the neighborhood 
of spatial infinity goes to $\pm{\delta_\mu}^\alpha$ as one approaches 
spatial infinity. In this case the total mass expressed in terms 
of the Komar integral can be rewritten as
\begin{eqnarray}
M &=&
\frac1{2\Omega_{d-2}}\oint_{{\cal S}_\infty}d{\cal S}^{\mu_1\ldots\mu_{d-2}}
 \epsilon_{\mu_1\ldots\mu_{d-2}\nu\sigma}
 \nabla^\nu\tau^\sigma \nonumber \\
  &=&
\frac1{2\Omega_{d-2}}\oint_{{\cal S}_\infty}d{\cal S}~e
 N^\mu\xi_\nu\nabla_\mu\tau^\nu \nonumber \\
  &=&
\frac1{2\Omega_{d-2}}\oint_{{\cal S}_\infty}d{\cal S}~
\frac12N^\mu\partial_\mu g_{tt}\nonumber \\
  &=&
\frac1{2\Omega_{d-2}}\oint_{{\cal S}_\infty}d{\cal S}~
N^\mu\left(\frac p r - A_t\right)F_{\mu t}\nonumber \\
  &=&
\frac {|Q|}2, \label{BPS}
\end{eqnarray}
which saturates the BPS bound $M\ge c|Q|$ with $c=1/2$.
$\tau^\mu=\delta^\mu_t$ is the timelike Killing vector and 
$Q$ is electric charge. 
All the fields are assumed to be $t$-independent 
in the neighborhood of infinity, in which the whole closed hypersurface 
${\cal S}_\infty$ ``infinitely close to infinity'' lies\footnote{
Strictly speaking,  
one must have a precise notion of asymptotic flatness to describe 
the limiting procedure more rigorously. This would be obtained by 
generalizing the definition in $d=4$ \cite{Ash80}. }. 
$N^\mu$ and $\xi^\mu$ are the unit future 
and outward pointing vectors normal to ${\cal S}_\infty$, respectively, 
which are orthogonal to each other. In the last line we used 
\eref{A-Relation} and ${e_t}^0\rightarrow\pm 1$. The fact that 
$c=1/2$ for $\chi=1$ can be confirmed by an explicit calculation  
using known BPS solutions such as extremal black holes.
It should be emphasized that the formula \eref{BPS} is derived from  
purely geometrical assumptions without any use of the field equations.

Because of the existence of extra isometries, $p$-brane solutions 
of usual type ($p\ge 1$) cannot be asymptotically Minkowskian in the 
$d$-dimensional sense. Therefore they are excluded from our discussion 
here (but can be easily discussed in parallel by considering mass/charge 
per unit world volume). Consequently magnetic charge is zero if $d\ge 5$. 
In $d=4$ it is also zero in our case since the way we identified the KK 
vector field implicitly assumed that there exists a smooth section of 
the $U(1)$ bundle (whose fiber is the orbit of 
$\partial/\partial x$-isometry) in the neighborhood of infinity.

The notion of BPS saturation is independent of how the dilaton is 
split from the $d$-metric (``frame''), but the constant factor 
$c$ changes.
For general $\chi$ the relation \eref{A-Relation} is replaced by 
\begin{equation}
{e_t}^0= \rho^{1-\chi}\left(\frac p r - A_t\right).
\label{generalA-Relation} \end{equation}
The expression of the total mass then depends also on the asymptotic 
behavior of the dilaton, which is not determined solely by geometrical  
constraints. The calculation using extremal black holes shows that 
the formula \eref{BPS} must be replaced by 
\begin{equation}
M=\frac{1+\chi}4 |Q|.
\label{BPS2}
\end{equation}

Let us now examine how the relation \eref{BPS2} is reconciled with 
the field equations in the case of the canonical metric 
$\chi=-\frac1{d-2}$. Dimensional
reduction of the Einstein--Hilbert action from $d+1$ dimensions to $d$
dimensions produces (up to a surface term) the standard result
\begin{equation}
ER(E)=
e\left[ R(e)-\frac14\rho^{2\frac{d-1}{d-2}}F_{\mu\nu}F^{\mu\nu}
        -\frac{d-1}{d-2}\partial_\mu\ln\rho\partial^\mu\ln\rho \right].
\end{equation}
Here $E\equiv\det{E_M}^A$, $e\equiv\det{e_\mu}^\alpha$, $R$
denotes the Ricci scalar, $F_{\mu\nu} := 2\partial_{[\mu}A_{\nu]}$ is
the KK field strength, and $d$-dimensional indices
$\mu,\nu$ are raised and lowered with the metric $g_{\mu\nu}:=
{e_\mu}^\alpha e_{\nu\alpha}$. The dilaton field equation, Maxwell's
equations, and Einstein's equation derived from the action are
\begin{eqnarray}
&&\nabla_\mu\partial^\mu\ln\rho
-\frac14\rho^{2\frac{d-1}{d-2}}F^2
= 0, \label{Dilaton} \\
&&\nabla_\mu\left(\rho^{2\frac{d-1}{d-2}}F^{\mu\nu}\right)
= 0, \label{Maxwell} \\
&&R_{\mu\nu}
-\frac12\rho^{2\frac{d-1}{d-2}}F_{\mu\sigma}{F_\nu}^\sigma
+\frac1{4(d-2)}g_{\mu\nu}\rho^{2\frac{d-1}{d-2}}F^2
-\frac{d-1}{d-2}\partial_\mu\ln\rho\partial_\nu\ln\rho
= 0, \label{Einstein}
\end{eqnarray}
where $\nabla_\mu$ denotes the covariant derivative associated with
the Levi--Civita connection derived from $g_{\mu\nu}$.

To proceed, we assume that the $d$-dimensional spacetime is parameterized 
by a single global coordinate system and that the timelike Killing vector 
$\tau\equiv\tau^M\partial_M=\partial_t$ is orthogonal
to a hypersurface parameterized by $x^m$, $m=1,\ldots,d-1$. 
Since $R_{\mu\nu}\tau^\nu=-\nabla^\nu\nabla_\nu\tau_\mu$ 
for any Killing vector $\tau$ 
(see e.g. \cite{Wald84}), we get
\begin{eqnarray}
M &=& 
 \frac1{2\Omega_{d-2}}\int_{\Sigma}
dV^{\mu_1\ldots\mu_{d-1}}
 \epsilon_{\mu_1\ldots\mu_{d-1}\sigma}
 \nabla_\nu\nabla^\nu\tau^\sigma \nonumber \\
&=& 
-\frac1{2\Omega_{d-2}}\int_{\Sigma}
dV^{x^1\ldots x^{d-1}}
 \epsilon_{x^1\ldots x^{d-1}t}
 {R^t}_t \nonumber \\
&=&
-\frac1{2\Omega_{d-2}}\int_{\Sigma}
dVe\frac12\rho^{2\frac{d-1}{d-2}}\left(F_{tm}F^{tm}-\frac{1}{2(d-2)}F^2
\right) 
\nonumber \\
&=&
-\frac1{2\Omega_{d-2}}\int_{\Sigma}
dV\nabla_\mu\left[e\rho^{2\frac{d-1}{d-2}}\left(
\frac12\left(1-\frac1{d-2}\right)
A_t
F^{\mu t}
-\frac1{2(d-2)}A_mF^{\mu m} 
\right)\right],\nonumber \\
\label{mass}
\end{eqnarray}
where $\Sigma:={\rm Int}{\cal S}_\infty$.
The second term becomes a surface integral, which can be interpreted  
as the net current flow through ${\cal S}_\infty$ provided that 
the $A_m\rho^{2\frac{d-1}{d-2}}$ factor varies sufficiently slowly near 
infinity. So let us assume here that it drops out.   
The first term can be also written as a surface integral. 
At first sight $A_t$ can be shifted by an arbitrary integration 
constant. This ambiguity can be fixed by taking account 
of the missing source term in RHS of Maxwell's equation \eref{Maxwell}. 
Indeed, it is the equation only for the vacuum region outside the 
locations of point charge, without which we would get $Q=0$. So suppose 
for simplicity that we have a single delta-function singularity at 
$p$. Then it picks up the value of $A_t$ at $p$ when the third equality  
in \eref{mass}
is written into a surface integral. 
This enables us to obtain the expression 
\begin{eqnarray}
M&=&\frac14(A_t(p)-A_t({\infty}))\left(1-\frac{1}{d-2}\right)Q\nonumber\\
&=&\frac14(\rho^{-\frac{d-1}{d-2}}{e_t}^0({\infty})-
\rho^{-\frac{d-1}{d-2}}{e_t}^0(p))\left(1-\frac{1}{d-2}\right)Q,
\label{BPS3}
\end{eqnarray}
which does not have the ambiguity any more. 

For asymptotically Minkowskian solutions ${e_t}^0(\infty)=\pm 1$.
If $\rho(\infty)=1$ and $\rho^{\frac{d-1}{d-2}}$ blows up 
at least faster than ${e_t}^0$ at $p$ 
(as is the case for extremal black holes), 
\eref{BPS3} becomes $M=\frac14(1-\frac{1}{d-2})|Q|$, which is in 
agreement with \eref{BPS2} with $\chi=-\frac1{d-2}$.

As a final comment we note that the mass expressed as a Komar 
integral \eref{BPS} or \eref{BPS2} coincides with the ADM mass 
in $d=4$ \cite{AshHan78} but in general they are different. For 
extremal black holes they differ by a factor 
$2(1-\frac1{d-2})$.

We have shown that the BPS saturation for strictly stationary, 
asymptotically Minkowskian solutions of a null reduction 
is a purely geometrical consequence. It would be 
interesting to investigate whether the relation between null 
reduction and BPS saturation extends to general non-stationary cases. 
A comparison with the supersymmtric string waves of \cite{BKO92}, 
which only assume a covariantly constant null (Killing) vector, 
may give us a hint.

Let us conclude with a speculation how the hyperbolic Kac--Moody
algebra $E_{10}$ might give rise to a duality symmetry of M theory. As
was shown in \cite{Nico92,Mizo97}, to obtain an $E_{10}$ hidden
symmetry in the dimensional reduction of $d=11$ supergravity, the
final step from $d=2$ to $d=1$ must be a null reduction.  The order in
which the dimensional reductions are performed should not affect the
final symmetry group. Hence suppose in particular that we first
perform a null reduction from $d=11$ to $d=10$ and then reduce the
other nine spatial dimensions. In this case all the solutions that
satisfy the conditions we assumed in this paper are BPS saturated in
$d=10$. Upon compactification we could also find corresponding BPS
solutions in lower dimensions (e.g.\ by taking a periodic array of
solutions). So this would mean that $E_{10}$ is a purely stringy
symmetry in the sense that its action may be defined only on BPS
solutions.

\vskip 5mm
We thank H Nicolai and E Witten for helpful comments. The work of R W
G was supported by {\it Deutsche Forschungsgemeinschaft} under
Contract No.\ DFG Ge 963/1-1. S M is grateful to the Institute for
Advanced Study in Princeton, where part of this work was carried out,
for hospitality and support.
%

\begin{thebibliography}{1}

\bibitem{HorStr91}
Horowitz G~T and Strominger A
\newblock 1991
\newblock {\it Black strings and p-branes}
\newblock {\it Nucl. Phys.} {\bf B360} 197

\bibitem{RusTse97}
Russo J~G and Tseytlin A~A
\newblock 1997
\newblock {\it Waves, boosted branes and {B}{P}{S} states in {M} theory}
\newblock {\it Nucl. Phys.} {\bf B490} 121

\bibitem{JulNic95}
Julia B and Nicolai H
\newblock 1995
\newblock {\it Null-{K}illing vector dimensional reduction and {G}alilean
  geom\-et\-ro\-dyn\-amics}
\newblock {\it Nucl. Phys.} {\bf B439} 291

\bibitem{Ash80}
Ashtekar A 
\newblock 1980
\newblock {\it Asymptotic structure of the gravitational field 
at spacial infinity }
\newblock {in {\it General relativity and gravitation} } Vol.{\bf 2} 
\newblock {ed. Held A}
\newblock (New York: Plenum)

\bibitem{Wald84}
Wald R~M
\newblock 1984
\newblock {\it General Relativity}
\newblock (University of Chicago Press: Chicago)

\bibitem{AshHan78}
Ashtekar A and Hansen R O
\newblock 1978
\newblock {\it A unified treatment of null and spacial infinity 
in general relativity. I. 
Universal structure, asymptotic symmetries,
and conserved quantities at spatial infinity
}
\newblock {\it J. Math. Phys.} {\bf 19} 1542

\bibitem{BKO92}
Bergshoeff E~A, Kallosh R and Ortin T
\newblock 1992
\newblock {\it Supersymmetric string waves}
\newblock {\it Phys. Rev.} {\bf D47} 5444

\bibitem{Nico92}
Nicolai H
\newblock 1992
\newblock {\it A hyperbolic {L}ie algebra from supergravity}
\newblock {\it Phys. Lett.} {\bf B276} 333

\bibitem{Mizo97}
Mizoguchi S
\newblock 1997
\newblock {\it $E_{10}$ symmetry in one-dimensional supergravity}
\newblock Tokyo U. INS preprint INS-1191 hep-th/9703160 

\end{thebibliography}


%
\end{document}